\documentclass{ckm}                 % twocolumn proceedings style
\usepackage{amsmath}

\def\rmO{{\rm O}}

\def\proof{\noindent{\sl Proof:}\kern0.6em}

\def\dual{\mathstrut^*\kern-0.1em}

\def\lvec#1{\setbox0=\hbox{$#1$}
    \setbox1=\hbox{$\scriptstyle\leftarrow$}
    #1\kern-\wd0\smash{
    \raise\ht0\hbox{$\raise1pt\hbox{$\scriptstyle\leftarrow$}$}}
    \kern-\wd1\kern\wd0}
\def\rvec#1{\setbox0=\hbox{$#1$}
    \setbox1=\hbox{$\scriptstyle\rightarrow$}
    #1\kern-\wd0\smash{
    \raise\ht0\hbox{$\raise1pt\hbox{$\scriptstyle\rightarrow$}$}}
    \kern-\wd1\kern\wd0}

\def\nabstar#1{\nabla\kern-0.5pt\smash{\raise 4.5pt\hbox{$\ast$}}
               \kern-4.5pt_{#1}}

\def\drvstar#1{\partial\kern-0.5pt\smash{\raise 4.5pt\hbox{$\ast$}}
               \kern-5.0pt_{#1}}

\def\MeV{{\rm MeV}}
\def\GeV{{\rm GeV}}

\def\fm{{\rm fm}}

\def\rhoprime{\rho\kern1pt'}
\def\rhobar{\bar{\rho}}
\def\rhobarprime{\rhobar\kern1pt'}
\def\rhobartilde{\kern2pt\tilde{\kern-2pt\rhobar}}
\def\rhobartildeprime{\kern2pt\tilde{\kern-2pt\rhobar}\kern1pt'}

\def\zetabar{\bar{\zeta}}
\def\zetaprime{\zeta\kern1pt'}
\def\zetabarprime{\zetabar\kern1pt'}
\def\zetar{\zeta_{\raise-1pt\hbox{\sixrm R}}}
\def\zetabarr{\zetabar_{\raise-1pt\hbox{\sixrm R}}}

\def\phiimpr{\phi_{\kern0.5pt\hbox{\sixrm I}}}

\def\diracstar#1#2{
    \setbox0=\hbox{$\gamma$}\setbox1=\hbox{$\gamma_{#1}$}
    \gamma_{#1}\kern-\wd1\kern\wd0
    \smash{\raise4.5pt\hbox{$\scriptstyle#2$}}}

\def\ba{b_{\rm A}}

\def\fa{f_{\rm A}}

\def\f1{f_1}

\def\opprime#1{\setbox0=\hbox{${\cal O}$}\setbox1=\hbox{${\cal O}_{\rm #1}$}
    {\cal O}_{\rm #1}\kern-\wd1\kern\wd0
    \smash{\raise4.5pt\hbox{\kern1pt$\scriptstyle\prime$}}\kern1pt}

\def\ophatprime#1{\setbox0=\hbox{$\widehat{\cal O}$}
    \setbox1=\hbox{$\widehat{\cal O}_{\rm #1}$}
    \widehat{\cal O}_{\rm #1}\kern-\wd1\kern\wd0
    \smash{\raise4.5pt\hbox{\kern1pt$\scriptstyle\prime$}}\kern1pt}

\def\bopprime#1{\setbox0=\hbox{${\cal O}$}\setbox1=\hbox{${\cal O}_{\rm #1}$}
    {\cal L}_{\rm #1}\kern-\wd1\kern\wd0
    \smash{\raise4.5pt\hbox{\kern1pt$\scriptstyle\prime$}}\kern1pt}

\def\blagprime#1{\setbox0=\hbox{${\cal B}$}\setbox1=\hbox{${\cal B}_{#1}$}
    {\cal B}_{#1}\kern-\wd1\kern\wd0
    \smash{\raise5.2pt\hbox{\kern1pt$\scriptstyle\prime$}}\kern1pt}

\def\za{Z_{\rm A}}

\def\msbar{{\rm \overline{MS\kern-0.05em}\kern0.05em}}
\def\MSbar{{\rm \overline{MS\kern-0.05em}\kern0.05em}}

\def\Ds{\rm D_s}
\def\fDs{F_{\Ds}}

\confname{Workshop on the CKM Unitarity Triangle, IPPP Durham, April
  2003}

\title{Precision computation of the leptonic $\Ds$-meson decay
  constant in quenched QCD}

\author{A J\"uttner and J Rolf\thanks{This work is part of the ALPHA
    research programme. JR thanks the organizers of the conference for
    the kind invitation.}\thanks{HU-EP-03/26, SFB/CPP-03-09}}
\address{Department of Physics, Humboldt-University Berlin}

\begin{document}

\begin{abstract}
We summarize a computation of the leptonic decay constant $\fDs$ of
the $\Ds$-meson in quenched QCD on
the lattice. We perform a direct simulation at the masses of the
strange and the charm quarks at four different lattice spacings from
approximately $0.1\,\fm$ to $0.05\,\fm$. Fully non perturbative
$\rmO(a)$-improvement is employed.
After taking the continuum limit we arrive at a value of
$\fDs=252(9)\,\MeV$, when setting the scale with the Kaon decay
constant $F_{\rm K}=160\,\MeV$. Setting the scale with the nucleon
mass instead leads to a decrease of about $20\,\MeV$ of $\fDs$. 
\end{abstract}

%% \maketitle needs to be after the author and address info and the
%% abstract... 
\maketitle

%% standard LaTeX from here on...

\section{Introduction}

To get reliable estimates of weak decay constants like $F_{\rm B}$
lattice QCD has often to be supplemented by chiral extrapolations
and/or heavy quark effective theory. These introduce substantial
systematic errors~\cite{ryan}. In addition, the usual
lattice errors like statistical errors, discretization errors, finite
volume effects, contamination from excited states, perturbative
renormalization and quenching~\cite{flynn} have to be
understood or eliminated. 

The $\Ds$-meson however is special in this
context. It consists of a strange and a charm quark. Both can be
implemented precisely directly on the lattice as has been done
recently by the ALPHA collaboration~\cite{smass,cmass}. Thus neither a
chiral extrapolation nor an extrapolation to heavy quarks (or similar
strategies) have to be used. Still we expect the $\Ds$-meson to be
similar to other heavy light systems. Therefore it can be studied to
understand all the other error sources mentioned above. 

The goal of our work, which has been recently published in~\cite{fds},
is a precision computation of $\fDs$ in quenched QCD. We aim at a
combined error of three percent. To eliminate the
discretization errors, we perform fully non perturbative
$\rmO(a)$-improvement and simulate at four different lattice spacings.
This allows us to take a reliable continuum limit. Finite volume
effects have been shown to be negligible in~\cite{smass}. At our masses
they are expected to be even smaller.  We {\sl define} our plateaus
such that the contamination by excited states stays below five per
mille. Finally all the uncertainty due to perturbative renormalization
has been eliminated by using the non perturbative renormalization
techniques of the ALPHA
collaboration~\cite{Jansen:1995ck,Luscher:1996jn}. The only 
systematic uncertainty we cannot deal with at present is the
quenching error.

A precise value of $\fDs$ in quenched QCD is desirable since together
with a computation of the decay constant in the static approximation
it will enable us to see how far the heavy quark effective theory can
be applied safely. In unquenched simulations we can then rely on this
experience in the computation of $F_{\rm B}$. 

The $\Ds$-meson is stable in QCD. It decays weakly by an emission of a
W-boson into a lepton and a neutrino. The branching ratios can be
measured experimentally. They are summarized in~\cite{pdb}. Given the
CKM-matrix element $V_{\rm cs}$, $\fDs$ can then be determined, since
the standard model predicts
\begin{eqnarray}
  &&\text{BR}(\Ds\rightarrow\text{l}\nu) =\\
  &&\frac{\text{G}_{\text{F}}^2}{8\pi}\ \tau_{\Ds}\
  {F_{\Ds}}^2\ 
  \vert V_{\text{cs}} \vert^2\ m_{\Ds}\ m_{\text{l}}^2 \left( 1-
    \frac{m_{\text{l}}^2}{m_{\Ds}^2} \right)^2.\nonumber
\end{eqnarray}
Recent
experimental data are shown in table~\ref{tab:exp}~\cite{exp}.
\begin{table}[htbp]
  \centering
  \begin{tabular}[c]{cc}
            & $F_{\Ds} [\MeV]$\\ \hline
            ALEPH & $285\pm 19 \pm 40$ \\
            DELPHI & $330\pm 95$\\
            L3 & $309\pm 58\pm 50$\\
            CLEO & $280 \pm 17\pm 42$\\
            BEATRICE & $323\pm 44\pm 36$\\
            E653 & $194\pm 35\pm 24$
  \end{tabular}
  \caption{Experimental data for $\fDs$.}
  \label{tab:exp}
\end{table}
The status of lattice computations was reviewed in~\cite{lat2} with a
quenched world average of $\fDs = 230\pm 14\,\MeV$. Table~\ref{tab:lattice}
summarizes some of the latest results. 
\begin{table}[htbp]
  \centering
  \begin{tabular}[c]{cc}
    & $F_{\Ds} [\MeV]$\\ \hline
    UKQCD~\cite{ukqcd} & $229(3)+23-12$ \\
    APE~\cite{ape} & $234(9)+5-0$ \\
    MILC~\cite{milc} & $223(5)+19-17$
  \end{tabular}
  \caption{Lattice results for $\fDs$ in quenched QCD.}
  \label{tab:lattice}
\end{table}
%With QCD sum rules one
%currently gets $\fDs = 235\pm 24\,\MeV$~\cite{sum1,sum2}.

\section{Strategy}

The decay constant $\fDs$ is defined by the QCD matrix element
\begin{equation}
  \langle 0\vert A_{\mu}(0)\vert \Ds(p)\rangle = i p_{\mu} \fDs
\end{equation}
of the axial current $A_{\mu} = \bar{\rm s}\gamma_{\mu}\gamma_5{\rm
  c}$.  To formulate this problem on the lattice we eliminate the bare
parameters of the QCD Lagragian in favour of physical observables in
one chosen hadronic scheme. Our strategy~\cite{npmass,smass,cmass} is
to use the kaon decay constant $F_{\rm K}$ to set the scale, that
means to compute the lattice spacing $a$ in physical units as a
function of the bare coupling $g_0$. The bare strange quark mass and
the bare charm quark mass are eliminated by the masses $m_K$ and
$m_{\Ds}$ of the kaon and the $\Ds$-meson, respectively. We neglect
isospin breaking and take the quark mass ratio $M_{\rm s}/M_{\rm
  light} = 24.4\pm 1.5$ from chiral perturbation theory~\cite{chiral}.
Here $M_{\rm light} = \frac{1}{2}(M_{\rm u} + M_{\rm d})$. The capital
letters denote the renormalization group invariant quark masses, which
are somewhat larger than the $\MSbar$-masses. In quenched QCD they
take the values $M_{\rm s} = 138(5)\,\MeV$~\cite{smass} and $M_{\rm c}
= 1.654(45)\,\GeV$~\cite{cmass}. The results mentioned above have been
obtained in units of the Sommer scale $r_0$ which is derived from the
force between static color sources. $r_0/a$ has been computed to a
high precision in~\cite{scale1,scale2} as a function of the bare
couling.  The Sommer scale is affected by lattice artifacts only at
$\rmO(a^2)$ in the quenched theory. Setting the scale with $F_{\rm K}
= 160\,\MeV$ is equivalent to using $r_0 = 0.5\,\fm$ while the nucleon
mass $m_N = 938\,\MeV$ roughly corresponds to
$r_0=0.55\,\fm$~\cite{smass}. This deviation is the typical size of
the quenched scale amiguity.

\section{Theory}

We perform numerical simulations in $\rmO(a)$-improved lattice QCD
using Schr\"odinger functional boundary conditions~\cite{nara,sint} on a
$L^3\times T$ space-time cylinder. For unexplained notation we refer
to~\cite{notation}. We define the
meson sources
\begin{equation}
  {\cal O} = a^6\sum_{{\bf y}, {\bf z}}\overline{\zeta}_j({\bf
    y})\gamma_5\zeta_i({\bf z}),
\end{equation}
\begin{equation}
  {\cal O}' = a^6\sum_{{\bf y}, {\bf z}}\overline{\zeta}'_j({\bf
    y})\gamma_5\zeta'_i({\bf z}), 
\end{equation}
with flavour indices $i$ and $j$ at the $x_0=0$ and $x_0=T$ boundary,
respectively. From these we compute the correlation functions
\begin{eqnarray}
  \label{eq:fA}
  f_{\rm A}^I(x_0) &=& -\frac{1}{2}\langle {\cal O} A_0(x)\rangle,\\
%  f_{\rm P}(x_0) &=& -\frac{1}{2}\langle {\cal O} P(x)\rangle,\\
  f_1 &=& - \frac{1}{3L^6} \langle {\cal O}' {\cal O}\rangle.
\end{eqnarray}
Here $A_{\mu}(x)$ denotes the improved axial current. It receives a
scale independent multiplicative renormalization $\za$ on the lattice.
In terms of these correlation functions $\fDs$ can be written as 
\begin{eqnarray}
\label{eq:fds}
  \fDs &=& -2 \za (1+\ba (am_{{\rm q},i}+am_{{\rm q},j})/2 )\nonumber\\
  &&\times\frac{\fa^I(x_0)}{\sqrt{f_1}} \
  (m_{\Ds}L^3)^{-1/2} e^{(x_0-T/2) m_{\Ds}} \nonumber\\
 &&\times
  \left\lbrace 1 - \eta_A^{\Ds} e^{-x_0\Delta} - \eta_A^0
    e^{-(T-x_0) m_{\text{G}}}\right\rbrace\nonumber\\
&&+ \rmO(a^2).
\end{eqnarray}
Here the factor $(m_{\Ds}L^3)^{-1/2}$ takes into account the
normalization of one particle states.  The contribution $f_1^{-1/2}$
cancels out the dependence on the meson sources.  Because of the
exponential decay of the correlation function $\fa^I$ the product
in~(\ref{eq:fds}) is expected to exhibit a plateau at intermediate
times when the contribution $\eta_A^{\Ds} e^{-x_0\Delta}$ of the first
excited state and the contribution $\eta_A^0 e^{-(T-x_0)
  m_{\text{G}}}$ from the $O^{++}$ glueball both are small.  A plateau
average can then be performed to increase the signal and is understood
in~(\ref{eq:fds}).  Further explanations for equation~(\ref{eq:fds})
and details can be found in~\cite{Guagnelli:1999zf}.

\section{Parameters}

We discretize the space-time cylinder using four different lattice
spacings $a$ but keep $L$ and $T=2L$ approximately constant in physical
units. To this end we use the same bare couplings that have been used
in the determinations of the strange and the charm quark masses
in~\cite{smass,cmass} by the ALPHA collaboration. From this work we
also take the hopping parameters for the quarks. Our choice of
parameters is shown in table~\ref{tab:par}.
\begin{table}[htbp]
  \begin{center}
    \begin{tabular}{lllll}\hline
      $\beta$ & $n_{\text{meas}}$ & $L/a$ & $\kappa_s$ & $\kappa_c$\\ \hline 
      6.0     & 380               & 16    & 0.133929 &  0.119053   \\
      6.1     & 301               & 24    & 0.134439 &  0.122490   \\
      6.2     & 251               & 24    & 0.134832 &  0.124637   \\
      6.45    & 289               & 32    & 0.135124 &  0.128131   \\ \hline
    \end{tabular}
    \caption{Statistics and parameters for our simulations.}
    \label{tab:par}
  \end{center}
\end{table}
In table~\ref{tab:const} we show that with this choice indeed the
physical conditions are constant to a sufficient precision.
\begin{table}[htbp]
  \begin{center}
    \begin{tabular}{lll}\hline
      $\beta$ & $L/r_0$ & $r_0 m_{\Ds}$ \\ \hline 
      6.0     &   2.98  & 4.972(22)     \\
      6.1     &   3.79  & 4.981(23)     \\
      6.2     &   3.26  & 5.000(25)     \\
      6.45    &   3.06  & 5.042(29)     \\ \hline
    \end{tabular}
    \caption{Demonstration of constant physical conditions.}
    \label{tab:const}
  \end{center}
\end{table}

\section{Computation of the decay constant}

\begin{figure}
\hbox to\hsize{\hss
\includegraphics[width=\hsize]{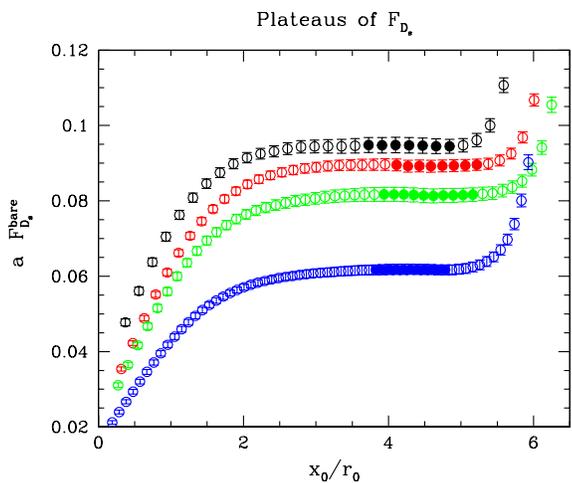}
\hss}
\caption{Behaviour of $a \fDs^{\text{bare}}$. The full symbols denote
  the plateau range.}
\label{fig:plateau}
\end{figure}

To compute the decay constant $\fDs$ we use the
combination~(\ref{eq:fds}) of correlation functions. For all parameter
choices we find plateaus as functions of $x_0$. These plateaus are
shown in figure~\ref{fig:plateau}. 
Their extent is roughly from $4r_0$ to $5r_0$. At small respectively
large times we fit to the functions $\fDs(x_0)$ the expected
contributions of the first excited state and the
glueball~(\ref{eq:fds}). For details see~\cite{fds}. We define the
plateau such that their sum stays below 5 per mille.

Since we deal with heavy quark propagators on APE1000 in single precision
we have to check that the rounding errors are small enough.
Our check against runs with a double precision code~\cite{milccode}
reveals that the rounding errors are smaller than one per mille.

\begin{figure}
\hbox to\hsize{\hss
\includegraphics[width=\hsize]{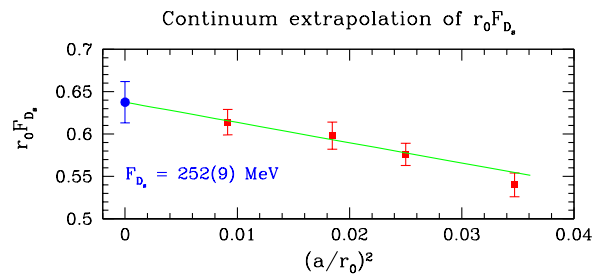}
\hss}
\caption{Continuum extrapolation of $\fDs$.}
\label{fig:cont}
\end{figure}

After averaging $\fDs$ over the plateaus defined above we get the
values shown in table~\ref{tab:fds}.
\begin{table}[htbp]
  \begin{center}
    \begin{tabular}{ll}\hline
      $\beta$ & $r_0 F_{\Ds}$ \\ \hline 
      6.0 & 0.540(14) \\
      6.1 & 0.576(13) \\
      6.2 & 0.598(16) \\
      6.45& 0.614(15) \\ \hline
      c.l.& 0.638(24) \\ \hline
    \end{tabular}
    \caption{Simulations results and continuum limit for $\fDs$.}
    \label{tab:fds}
  \end{center}
\end{table}
These data can be extrapolated to the continuum limit. Here we leave out
the coarsest lattice. The
extrapolation can be performed in $(a/r_0)^2$ since we employ non
perturbative $\rmO(a)$ improvement. Here we take $\ba$ from the Los
Alamos group~\cite{ba}. Since this involves an extrapolation of their
data we have also used 1-loop perturbation theory~\cite{baPT}. Then we get
$r_0\fDs = 0.631(24)$, which is in perfect agreement with our main
result, $r_0\fDs=0.638(24)$, or, using $r_0 = 0.5\,\fm$, $\fDs =
252(9)\,\MeV$. The continuum extrapolation is shown in
figure~\ref{fig:cont}. 

\section{The quenched scale ambiguity}

To estimate the quenched scale ambiguity of $\fDs$ under a scale shift
of 10 percent, which is typical for the quenched approximation, we
consider $r_0\fDs = f(z)$ as a function of the meson mass $z = r_0
m_{\Ds}$. We expand $f(z)$ around the physical value $z_0=4.988$ up to
first order. A 10 percent increase of $r_0$ corresponds to
$z-z_0=0.5$. With an estimate of $f'(z_0)$ from a linear fit of
$r_0\fDs$ around $m_{\Ds}$ we get $f(z)-f(z_0)\approx
0.008(3)$. Converting back to physical units, now using
$r_0=0.55\,\fm$ we find that $\fDs$ decreases by $20\,\MeV$,
corresponding to eight percent. The estimate of $f'(z_0)$ is possible
since in addition to the hopping parameters already discussed we have
also performed simulations around the charm quark mass value.

\section{Conclusion}

The leptonic $\Ds$-meson decays can be studied on the lattice without
chiral extrapolations or heavy quark effective theory. This has
enabled us to perform a computation of $\fDs$ with a precision that
matches the precision  goals at future experiments, for example,
CLEO. The precise value of $\fDs$ in quenched QCD, together with new
precise data in the static approximation, will show how far heavy
quark effective theory can be applied safely. This is of importance
for the unquenched computation of $F_{\rm B}$ in the future.

We will supplement this analysis with more data around the charm
mass. Part of this data has already been used to estimate the quenched
scale ambiguity of $\fDs$ under a scale shift of 10 percent. 

\section*{Acknowledgements}

This work was supported by the European Community under the grant
HPRN-CT-2000-00145 Hadrons/Lattice QCD and by the Deutsche
Forschungsgemeinschaft in the SFB/TR 09 and the Graduiertenkolleg
GK271. All the production runs were carried out on machines of the
APE1000 series at DESY. 
The check of the rounding errors ran
on the IBM p690 system of the HLRN (http://www.hlrn.de) and on the PC
cluster at DESY. We thank the staff at the computer centres
for their help and M. Della Morte, S. Sint, R. Sommer, H. Wittig and
U. Wolff for useful discussions.


\begin{thebibliography}{9}

%\cite{Kronfeld:2002ab}
\bibitem{ryan}
A.~S.~Kronfeld and S.~M.~Ryan,
%``Remark on the theoretical uncertainty in B0 anti-B0 mixing,''
Phys.\ Lett.\ B {\bf 543} (2002) 59.
%[arXiv:hep-ph/0206058].
%%CITATION = HEP-PH 0206058;%%


%\cite{Flynn:1997ca}
\bibitem{flynn}
J.~M.~Flynn and C.~T.~Sachrajda,
%``Heavy quark physics from lattice QCD,''
Adv.\ Ser.\ Direct.\ High Energy Phys.\  {\bf 15} (1998) 402.
%[arXiv:hep-lat/9710057].
%%CITATION = HEP-LAT 9710057;%%



%\cite{Garden:2000fg}
\bibitem{smass}
J.~Garden, J.~Heitger, R.~Sommer and H.~Wittig  [ALPHA and UKQCD Collaboration],
%``Precision computation of the strange quark's mass in quenched QCD,''
Nucl.\ Phys.\ B {\bf 571} (2000) 237.
%[arXiv:hep-lat/9906013].
%%CITATION = HEP-LAT 9906013;%%

%\cite{charm}
\bibitem{cmass}
J.~Rolf and  S.~Sint [ALPHA Collaboration],
%``A precise determination of the charm quark's mass in quenched QCD,''
JHEP 12 (2002) 007.
%[arXiv:hep-ph/0209255].

%\cite{fds}
\bibitem{fds}
A.~J\"uttner and J.~Rolf [ALPHA Collaboration],
%``A precise determination of the decay constant of the $\Ds$-meson in
%quenched QCD,''
Phys.\ Lett.\ B {\bf 560} (2003) 59.

%\cite{Jansen:1995ck}
\bibitem{Jansen:1995ck}
K.~Jansen {\it et al.},
%``Non-perturbative renormalization of lattice QCD at all scales,''
Phys.\ Lett.\ B {\bf 372} (1996) 275
[arXiv:hep-lat/9512009].
%%CITATION = HEP-LAT 9512009;%%


%\cite{Luscher:1996jn}
\bibitem{Luscher:1996jn}
M.~L\"uscher, S.~Sint, R.~Sommer and H.~Wittig,
%``Non-perturbative determination of the axial current normalization  constant in O(a) improved lattice QCD,''
Nucl.\ Phys.\ B {\bf 491} (1997) 344.
%[arXiv:hep-lat/9611015].
%%CITATION = HEP-LAT 9611015;%%


\bibitem{pdb}
K.~Hagiwara {et al.} [Particle Data Group Collaboration],
Phys.\ Rev.\ D{\bf 66} (2002) 010001.


%\cite{expsummary}
\bibitem{exp}
S.~S\"{o}ldner-Rembold,
%``Experimental Status of Leptonic $\Ds$ Decays'',
JHEP Proceedings, HEP2001, hep-ex/0109023.

%\cite{Draper:1998ms}
%\bibitem{lat1}
%T.~Draper,
%%``Status of heavy quark physics on the lattice,''
%Nucl.\ Phys.\ Proc.\ Suppl.\  {\bf 73} (1999) 43.
%%[arXiv:hep-lat/9810065].
%%%CITATION = HEP-LAT 9810065;%%

%\cite{Ryan:2001ej}
\bibitem{lat2}
S.~M.~Ryan,
%``Heavy quark physics from lattice QCD,''
Nucl.\ Phys.\ Proc.\ Suppl.\  {\bf 106} (2002) 86.
%[arXiv:hep-lat/0111010].
%%CITATION = HEP-LAT 0111010;%%

%%\cite{Bernard:2000ki}
%\bibitem{lat2}
%C.~W.~Bernard,
%%``Heavy quark physics on the lattice,''
%Nucl.\ Phys.\ Proc.\ Suppl.\  {\bf 94} (2001) 159.
%%[arXiv:hep-lat/0011064].
%%%CITATION = HEP-LAT 0011064;%%


%\cite{Maynard:2001zd}
\bibitem{ukqcd}
C.~M.~Maynard  [UKQCD Collaboration],
%``Heavy-light decay constants on the lattice,''
Nucl.\ Phys.\ Proc.\ Suppl.\  {\bf 106} (2002) 388.
%[arXiv:hep-lat/0109026].
%%CITATION = HEP-LAT 0109026;%%

%\cite{Becirevic:2000kq}
\bibitem{ape}
D.~Becirevic,
%``Heavy quark phenomenology from lattice QCD,''
Nucl.\ Phys.\ Proc.\ Suppl.\  {\bf 94} (2001) 337.
%[arXiv:hep-lat/0011075].
%%CITATION = HEP-LAT 0011075;%%

%\cite{Bernard:2000ht}
\bibitem{milc}
C.~W.~Bernard {\it et al.},
%``f(B) for various actions: Approaching the continuum limit with  dynamical fermions,''
Nucl.\ Phys.\ Proc.\ Suppl.\  {\bf 94} (2001) 346.
%[arXiv:hep-lat/0011029].
%%CITATION = HEP-LAT 0011029;%%

%%\cite{Narison:2001pu}
%\bibitem{sum1}
%S.~Narison,
%%``c, b quark masses and f(D/s), f(B/s) decay constants  from pseudoscalar sum rules in full QCD to order alpha(s)**2,''
%Phys.\ Lett.\ B {\bf 520} (2001) 115.
%%[arXiv:hep-ph/0108242].
%%%CITATION = HEP-PH 0108242;%%
%
%
%%\cite{Penin:2001ux}
%\bibitem{sum2}
%A.~A.~Penin and M.~Steinhauser,
%%``Heavy-light meson decay constant from QCD sum rules in three-loop  approximation,''
%Phys.\ Rev.\ D {\bf 65} (2002) 054006.
%%[arXiv:hep-ph/0108110].
%%%CITATION = HEP-PH 0108110;%%

%\cite{Capitani:1999mq}
\bibitem{npmass}
S.~Capitani, M.~L\"uscher, R.~Sommer and H.~Wittig  [ALPHA Collaboration],
%``Non-perturbative quark mass renormalization in quenched lattice QCD,''
Nucl.\ Phys.\ B {\bf 544} (1999) 669.
%[arXiv:hep-lat/9810063].
%%CITATION = HEP-LAT 9810063;%%

%\cite{Leutwyler:1996qg}
\bibitem{chiral}
H.~Leutwyler,
%``The ratios of the light quark masses,''
Phys.\ Lett.\ B {\bf 378} (1996) 313.
%[arXiv:hep-ph/9602366].
%%CITATION = HEP-PH 9602366;%%

%\cite{Guagnelli:1998ud}
\bibitem{scale1}
M.~Guagnelli, R.~Sommer and H.~Wittig, % [ALPHA collaboration],
%``Precision computation of a low-energy reference scale in
% quenched  lattice QCD,''
Nucl.\ Phys.\ B {\bf 535} (1998) 389.
%[arXiv:hep-lat/9806005].
%%CITATION = HEP-LAT 9806005;%%
%
%
%\cite{Necco:2001xg}
\bibitem{scale2}
S.~Necco and R.~Sommer,
%``The N(f) = 0 heavy quark potential from short to intermediate distances,''
Nucl.\ Phys.\ B {\bf 622} (2002) 328.
%[arXiv:hep-lat/0108008].
%%CITATION = HEP-LAT 0108008;%%


%\cite{Luscher:1992an}
\bibitem{nara}
M.~L\"uscher, R.~Narayanan, P.~Weisz and U.~Wolff,
%``The Schrodinger functional: A Renormalizable probe for nonAbelian gauge theories,''
Nucl.\ Phys.\ B {\bf 384} (1992) 168.
%[arXiv:hep-lat/9207009].
%%CITATION = HEP-LAT 9207009;%%

% SF for QCD

%\cite{Sint:1993un}
\bibitem{sint}
S.~Sint,
%``On the Schrodinger functional in QCD,''
Nucl.\ Phys.\ B {\bf 421} (1994) 135.
%[arXiv:hep-lat/9312079].
%%CITATION = HEP-LAT 9312079;%%

%\cite{Luscher:1996sc}
\bibitem{notation}
M.~L\"uscher, S.~Sint, R.~Sommer and P.~Weisz,
%``Chiral symmetry and O(a) improvement in lattice QCD,''
Nucl.\ Phys.\ B {\bf 478} (1996) 365.
%[arXiv:hep-lat/9605038].
%%CITATION = HEP-LAT 9605038;%%

%\cite{Guagnelli:1999zf}
\bibitem{Guagnelli:1999zf}
M.~Guagnelli, J.~Heitger, R.~Sommer and H.~Wittig  [ALPHA Collaboration],
%``Hadron masses and matrix elements from the QCD Schroedinger functional,''
Nucl.\ Phys.\ B {\bf 560} (1999) 465.
%[arXiv:hep-lat/9903040].
%%CITATION = HEP-LAT 9903040;%%

\bibitem{milccode}
We have used a modified
version of the public lattice gauge theory code of the MILC collaboration.
{\scriptsize http://www.physics.indiana.edu/\~{}sg/milc.html.}


%\cite{Bhattacharya:2001ks}
\bibitem{ba}
T.~Bhattacharya, R.~Gupta, W.~j.~Lee and S.~Sharpe,
%``Scaling behavior of improvement and renormalization constants,''
Nucl.\ Phys.\ Proc.\ Suppl.\  {\bf 106} (2002) 789.
%[arXiv:hep-lat/0111001].
%%CITATION = HEP-LAT 0111001;%%


%\cite{Guagnelli:2000jw}
%\bibitem{}
%M.~Guagnelli, R.~Petronzio, J.~Rolf, S.~Sint, R.~Sommer
%and U.~Wolff  [ALPHA Collaboration],
%``Non-perturbative results for the coefficients b(m) and b(A)-b(P) in
%O(a) improved lattice QCD,'' 
%Nucl.\ Phys.\ B {\bf 595} (2001) 44.
%[arXiv:hep-lat/0009021].
%%CITATION = HEP-LAT 0009021;%%

%\cite{Sint:1997jx}
\bibitem{baPT}
S.~Sint and P.~Weisz,
%``Further results on O(a) improved lattice QCD to one-loop order of  perturbation theory,''
Nucl.\ Phys.\ B {\bf 502} (1997) 251.
%[arXiv:hep-lat/9704001].
%%CITATION = HEP-LAT 9704001;%%
\end{thebibliography}
\end{document}